\documentclass[pra,amssymb,amsmath,superscriptaddress,twocolumn]{revtex4}

\usepackage{color}
\newtheorem{thm}{Theorem}

\newcommand{\be}{\begin{equation}}
\newcommand{\ee}{\end{equation}}

\begin{document}

\title{Dual  quantum  information   splitting  with  degenerate  graph
  states}  \author{Akshata  Shenoy  H} 
\email{akshata@ece.iisc.ernet.in}  
\affiliation{Applied  Photonics
  Lab,       ECE       Dept.,       IISc,      Bangalore,       India}
       \author{R.         Srikanth}
\email{srik@poornaprajna.org}  \affiliation{Poornaprajna  Institute of
  Scientific  Research, Bengaluru, India}  \affiliation{Raman Research
  Institute,     Bengaluru,     India.}      \author{T.      Srinivas}
\affiliation{Applied Photonics Lab, ECE Dept., IISc, Bangalore, India}

\begin{abstract}
We  propose  a  protocol  for  secret  sharing,  called  dual  quantum
information  splitting (DQIS), that  reverses the  roles of  state and
channel in standard quantum  information splitting.  In this method, a
secret is shared  via teleportation of a fiducial  input state over an
entangled state  that encodes the secret  in a graph  state basis.  By
performing a  test of  violation of a  Bell inequality on  the encoded
state,  the   legitimate  parties   determine  if  the   violation  is
sufficiently high  to permit distilling  secret bits.  Thus,  the code
space must  be maximally  and exclusively nonlocal.   To this  end, we
propose two ways to obtain code words that are degenerate with respect
to  a Bell  operator.  The  security of  DQIS comes  from  monogamy of
nonlocal  correlations,  which we  illustrate  by  means  of a  simple
single-qubit  attack model.   The nonlocal  basis of  security  of our
protocol makes it suitable for security in general monogamous theories
and in the more stringent, device-independent cryptographic scenario.
\end{abstract}

\maketitle


\section{Introduction}

Quantum entanglement  enables tasks in  communication and cryptography
not  possible  in the  classical  world,  e.g., quantum  teleportation
\cite{qtele}, dense coding and unconditionally secure key distribution
\cite{rmpqkd}.   Experimental  breakthroughs  have  enabled  practical
creation and manipulation of entanglement \cite{qexp1}, an achievement
duly  recognized  by  the  2012  Nobel  prizes  in  physics.   Several
teleportation-based protocols  with multi-particle channels  have been
proposed  \cite{qtele0, qtele1,  qtele2, qtele3,  jams,  zhe06, sre08,
  qssband,  qteleaa}.  In  particular,  entanglement can  be used  for
quantum secret sharing (QSS),  the quantum version of classical secret
sharing  \cite{gottqss}.   QSS  involves  a  secret  dealer  splitting
information,  representing the  secret  quantum state  $|\Psi\rangle$,
among a  number of agents, such  that only authorized  subsets of them
can reconstruct the secret.

A protocol  for splitting quantum  information, and teleporting  it to
more than one  party over an entangled channel, such  that a subset of
agents  sharing   the  entanglement,   is  able  to   reconstruct  the
information, was first proposed in Ref.  \cite{hbb99}, further studied
by  various  authors  \cite{qsshi,qssami,qssqecc,qssuds,qss2epr},  and
also  implemented  experimentally  \cite{qssexp,qssexp1,qssexp0}  (the
last employing  only sequential measurements  on a single  qubit).  We
will  refer to  such  teleportation-based QSS  as quantum  information
splitting (QIS).  Both  QSS and QIS can be used  to share both quantum
and classical secrets.

An important  resource of entanglement are  \textit{graph states} that
are useful  in quantum error correction  \cite{schli}, one-way quantum
computing  \cite{1way}  and cryptography  \cite{sre08,db08,kfms,aziz}.
They have  been studied  extensively theoretically, and  been realized
experimentally recently \cite{grafexp1,grafexp2}.

Given a graph $G = (V,E)$ defined  by the set $V$ of $n$ vertices, and
set $E$ of  edges, we denote by $\mathcal{N}(j)$,  the set of vertices
with   which    vertex   $j$   is   connected   by    an   edge   (the
neighborhood). Corresponding  to each vertex $j$, one  can associate a
stabilizer operator:
\begin{equation}
g_j = X_j\bigotimes_{k \in \mathcal{N}(j)}Z_k,
\label{eq:generator}
\end{equation}
where $Z_k$ and $X_k$, along  with $Y_k$, denote Pauli matrices acting
on qubit  $k$.  We define  the graph state  basis by the  $2^n$ common
eigenstates $|G_{\bf  x}\rangle \equiv |G_{x_1x_2\cdots  x_n}\rangle =
\bigotimes_j \left(Z_j\right)^{x_j}|G_{000\cdots0}\rangle$, with ($x_j
\in   \{0,1\}$)  of   the   $n$  commuting   operators  $g_j$,   where
$g_j|G_{x_1x_2\cdots    x_n}\rangle    =    (-1)^{x_j}|G_{x_1x_2\cdots
  x_n}\rangle$.   In particular, the  canonical $n$-qubit  graph state
$|G\rangle  \equiv  |G_{00\cdots0}\rangle$  is  characterized  by  $n$
independent perfect correlations of the form
\begin{equation}
g_j|G\rangle = |G\rangle.
\label{eq:graph}
\end{equation}
The  set  of all  $2^n$  products ($h_k$)  of  the  $g_j$'s forms  the
stabilizer group  $\mathcal{S}$. It follows  from Eq. (\ref{eq:graph})
that $h_j|G\rangle = |G\rangle$  for all $h_j \in \mathcal{S}$.  Graph
states are robust against decoherence \cite{grafdecoh}, which enhances
their practical value.

An  alternate equivalent definition  of graph  states, based  on their
generation via an Ising type of interaction, is as follows:
\begin{equation}
|G\rangle = \Pi_{(j,k) \in E} \mathcal{C}_Z^{\{j,k\}}|{+}\rangle,
\label{eq:ising}
\end{equation}
where $\mathcal{C}_Z$  is the controlled-phase gate.  Here  we use the
usual   notation  $Z|0\rangle   =   |0\rangle,  Z|1\rangle=-|1\rangle,
X|\pm\rangle = \pm|\pm\rangle$.

A special class of graph states are the linear cluster states, which
correspond to a linear graph. An $n$-qubit cluster state is given by:
\begin{equation}
|\phi_N\rangle =  \frac{1}{2^{n/2}} \bigotimes_j (|0\rangle +
|1\rangle_j  Z_{j+1}),
\label{eq:lc}
\end{equation}
with $Z_{n+1} \equiv 1$. For example,
\begin{equation}
|\phi_4\rangle =  \frac{1}{2}\left(|{+}0{+}0\rangle + |{+}0{-}1\rangle
+ |{-}1{-}0\rangle + |{-}1{+}1\rangle\right),
\label{eq:lc4}
\end{equation}
where     we     use      the     notation     $|{+}0{+}0\rangle     =
|{+}\rangle|0\rangle|{+}\rangle|0\rangle$,  etc.  It  should  be noted
that different  graphs may lead to  the same graph  state modulo local
transformations. For example, a star  graph over $n$ vertices leads to
the same state  irrespective of vertex it is  rooted in. The principal
graph transformation that leaves  the entanglement property of a graph
state invariant is \textit{local complementation} \cite{guhcab}.

As highly  entangled states,  graph states show  nonlocal correlations
\cite{sca05,genard,   bellgraph,   bellcab,   geza,  cabguh,   vinper,
  hsugraph, mard,  guhcab, bellmermin} that  contradict the assumption
of  local-realism, as  demonstrated  by their  violation of  Bell-type
inequalities  \cite{bell,chsh}.   This  is of  cryptographic  interest
because there is a close connection between security and the violation
of  a  Bell-type  inequality  \cite{ekert,sg1,sg2}.   This  connection
assumes further  importance in  the device independent  (DI) scenario,
where eavesdropper Eve is  allowed to conceal additional dimensions in
the devices of  legitimate parties, that empower a  side channel which
leaks basis and output information to Eve.

The  remaining   article  is   structured  as  follows.    In  Section
\ref{sec:dqis}, we  introduce a twist to  the QIS idea,  which we term
\textit{dual}  QIS, or  DQIS, wherein  a  fixed fiducial  state of  an
ancilla is teleported over an  entangled state that encodes the secret
and  satisfies  certain   conditions  of  teleportation.   In  Section
\ref{sec:nonlocal}, we  study the nonlocality of the  DQIS code space,
pointing out  two ways of contructing  Bell-type inequalities suitable
to witness its nonlocality.  In Section \ref{sec:security}, we discuss
the  security of  the DQIS  based on  the violation  of  the Bell-type
inequalities,  in  particular,  touching upon  the  device-independent
scenario  \cite{diqkd0,diqkd2}.  A  simple  single-qubit eavesdropping
attack  on DQIS  based on  a 5-qubit  1-bit error  correcting  code is
given,  to illustrate  how  Eve's entangling  action  can be  detected
because  of monogamy  of quantum  nonlocal correlations.   Finally, we
conclude in Section \ref{sec:conclusions}.

\section{Dual quantum information splitting \label{sec:dqis}}

In standard  QIS, one teleports  an unknown state  $|\Psi\rangle$ (the
secret) over  a teleportation channel,  which is a  suitably entangled
state.  By  contrast, in DQIS, we teleport  a \textit{fiducial} state,
$|0\rangle$ by convention,  across an entangled state $|\Psi_L\rangle$
that  encodes  $|\Psi\rangle$,  such   that  the  end  result  of  the
teleportation is the recovery of $|\Psi\rangle$.

DQIS can be useful in situations where the qudit secret $|\Psi\rangle$
is priorly known  to the dealer Alice, before  the distribution of the
entangled particles  to the agents,  and furthermore it is  unsafe for
Alice to store $|\Psi\rangle$ indefinitely in her station. This may be
the case  in situations  where Alice has  bounded quantum  memory, and
cannot stock secrets (in addition  to her entangled particles), but is
able  to prepare  a fixed  state when  transmission is  needed.  Alice
classically  encrypts  $|\Psi\rangle$ using  one  of $d^2$  operations
\cite{qssuds}, and transmits it to a distributor Dolly, who encodes it
into an entangled state, which is transmitted to all relevant parties.

For  example,  the  classical  encryption  of  a  qubit  requires  the
equi-probable  application  of  the  4 four  Pauli  operations,  which
transforms  a qubit  in  an  arbitrary state  into  a maximally  mixed
state. Alice  must divulge  the two-bit (in  general, $2\log  d$ bits)
decyrption information for recovery. Of  course the dealer may also be
the distributor.

The basic DQIS protocol works as follows:
\begin{enumerate}
\item  Alice prepares  the $N$  copies of  the  $d$-dimensional secret
  $|\Psi\rangle   =  \sum_{j=1}^d  \alpha_j   |j\rangle$,  classically
  encrypts each of them, and transmits them to Dolly, the distributor.
\item Dolly encodes each of them into an entangled state consisting of
  a superposition  of suitable graph basis states.  For example, state
  $|\Psi\rangle$ is encoded as:
\begin{equation}
|\Psi_L\rangle = \sum_{j=1}^d \alpha_j|G_j\rangle.
\label{eq:encode}
\end{equation}
The $|G_j\rangle$'s are  chosen so that they are  suitable for QIS and
satisfy an additional, teleportation condition discussed below.
\item Dolly transmits  the qubits in her possession  to the legitimate
  parties Alice,  Bob, Charlie, Rex,  et al.  After their  receipt has
  been  acknowledged  over  an  authenticated classical  channel,  she
  randomly  selects $N-1$ of  the transmitted  states ,  and announces
  their serial numbers.
\item The parties perform  their local operations chosen randomly from
  a pre-agreed set, and communicate their classical outputs to Alice.
\item Alice  performs a basis  reconciliation where she  determines if
  the  measurements  are appropriate  to  compute pre-agreed  products
  (stabilizers $h_j$)  of local Pauli operations on  the particles. If
  the measurements correspond to  none of the pre-agreed $h_j$'s, they
  are  discarded. Else,  they  are used  to  test the  violation of  a
  Bell-type inequality, which has the form:
  \begin{equation}  \langle
\mathcal{B} \rangle \equiv \sum_{j=1}^m \langle h_j \rangle \le 2q-m,
\label{eq:bell}
\end{equation}
where  $\mathcal{B}$ is the  Bell operator  and $q$  ($\le m$)  is the
largest number of  the $h_j$'s that assume a  positive value ($+1$) if
each particle  is assumed  to possess  a definite value  of $X,  Y, Z$
irrespective  of the  measurement setting  on any  other  particle.  A
contradiction with  \textit{local-realism}, and hence  a demonstration
of quantum nonlocality, occurs when $q<m$.

The  quantum  bound  on  the  l.h.s of  Eq.   (\ref{eq:bell})  is  the
algebraically allowed  maximum of $m$.  Alice determines  if the basis
reconciled  correlation   data  derived  from  the   $N-1$  states  is
compatible  with distillable secrecy  (by checking  if they  produce a
sufficiently high violation of a Bell inequality).
\item  If  the inequality  (\ref{eq:bell})  is  found  to be  violated
  sufficiently highly, Alice teleports the fiducial state $|0\rangle$,
  and  signals the  other agents  to proceed  to the  next  step: they
  perform  standard teleportation measurements  on their  particles on
  the unmeasured state, and convey the resulting classical information
  to the recoverer, Rex.
\item  Rex  recovers  the  encrypted  secret based  on  the  classical
  communication from all other parties.
\item Alice gives Rex the classical decryption information, from which
  Rex recovers $|\Psi\rangle$.
\end{enumerate}
The type of  encoding in Eq.  (\ref{eq:encode}) must  be such that Rex
recovers $|\Psi\rangle$  even though Alice  teleports $|0\rangle$. The
conditions  under which  this  works are  discussed  in the  following
Section \ref{sec:dqiscond}. It might  be thought that since the secret
is encoded  in the distributed entanglement, and  the teleported state
is  publicly  known,  therefore  the  teleportation  may  be  entirely
eliminated. Still, the teleportation is needed so that the distributed
state can be accessed via local operations and classical communication
(LOCC) between the parties.

The issue of conditions under which  the parties are be able perform a
test of Bell  inequality violation on the encoded  state, is discussed
in  Section  \ref{sec:nonlocal}.   From an  experimental  perspective,
implementing our protocol is not  expected to be difficult in a set-up
that realizes  graph states, since the only  additional requirement is
creation of superposition of these states.

\subsection{DQIS conditions for a qudit \label{sec:dqiscond}}

We  define  a \textit{teleportation  configuration}  \textbf{C} as  an
arrangement  of agents  and their  actions that  fixes who  the secret
dealer (Alice)  is, who the recoverer  (Rex) is, etc.,  and what their
local  operations  are.   Two  such  basis  states,  which  we  denote
$|G_0\rangle$   and  $|G_1\rangle$,  constitute   a  teleportationally
\textit{divergent} pair, if for a  fiducial input state (taken here to
be  $|0\rangle$),  Rex  recovers $|j\rangle$  ($|\overline{j}\rangle$)
when  the  channel is  $G_0$  ($G_1$),  for  a given  \textbf{C},  and
classical measurement outcome \textbf{M} of all other parties.

More generally, consider a $d$-dimensional secret (a qudit state), and
$n$-qubit  graph basis states  $|G_j\rangle$ ($j=1,\cdots,d$  where $d
\le 2^n$) associated with a graph $G(E, V)$.  Letting:
\begin{equation}
|\psi_j\rangle_R    =   {_{uA\xi}\langle}    \upsilon_k   |0\rangle_u|
G_j\rangle_{AR\xi},
\label{eq:dqiscond}
\end{equation}
teleportation  divergence   entails  that  there   exists  a  recovery
operation $U^\dag_{\bf C, M}$ such that:
\begin{equation}
|\psi_j\rangle =  U_{\bf C, M}|j\rangle.
\label{eq:complementary}
\end{equation}
Here the labels $u, A, R$ and $\xi$ denote the ancilla, Alice, Rex and
the remaining agents; $|\upsilon_k\rangle$ is a particular measurement
outcome collectively obtained by all but Rex.  The set of graph states
that satisfy (\ref{eq:dqiscond})  and (\ref{eq:complementary}) for all
$|\upsilon_k\rangle$   allowed  under  \textbf{C}   are  said   to  be
\textit{teleportation divergent}.

Eq.   (\ref{eq:dqiscond})   can  be  regarded  as   defining  the  map
$\mathcal{T}_{\bf  C,M}$, for  a given  measurement  configuration and
conditioned on measurement outcomes  \textbf{M}.  We have by virtue of
quantum   mechanical  linearity,  Eqs.   (\ref{eq:complementary})  and
(\ref{eq:encode})
\begin{equation}
\mathcal{T}_{\bf C,M}\left(|\Psi_L\rangle\right) 
= U_{\bf C,M}\left(\sum_j \alpha_j|j\rangle\right).
\label{eq:tele}
\end{equation}
Thus,  conditioned  on   the  classical  information  \textbf{M},  Rex
recovers the secret $|\Psi\rangle$.

In the following two subsections, we illustrate DQIS of a qubit secret
with a 4-qubit cluster state subspace, and the code space of a 5-qubit
quantum error  correcting (QEC) code. To conclude  this subsection, we
consider a simple example where  DQIS fails for a particular choice of
$|G_j\rangle$ and \textbf{C}. These are taken to be graph basis states
given by the GHZ class states
\begin{eqnarray}
|G_{000}\rangle &=& \frac{1}{\sqrt{2}}(|000\rangle + |111\rangle),
\nonumber \\
|G_{100}\rangle &=& \frac{1}{\sqrt{2}}(|000\rangle - |111\rangle),
\label{eq:ghz}
\end{eqnarray}
We choose the configuration where Alice holds qubit 1 and measures the
ancillary qubit and  her entangled qubit in the  Bell basis, Bob holds
qubit 2  and measures in the  $X$ basis, and Charlie  must recover the
secret.  One finds:
\begin{widetext}
\begin{eqnarray}
\alpha|0\rangle_C   +    \beta|1\rangle_C   &\propto&   {_{uA}\langle}
\Phi^+|{_B\langle}{+}|(\alpha|0\rangle_u                              +
\beta|1\rangle_u)|G_{000}\rangle_{ABC}, \nonumber \\ \alpha|0\rangle_C
- \beta|1\rangle_C               &\propto&              {_{uA}\langle}
\Phi^+|{_B\langle}{+}|(\alpha|0\rangle_u                              +
\beta|1\rangle_u)|G_{100}\rangle_{ABC}.
\label{eq:dqiscond0}
\end{eqnarray}
\end{widetext}
As the l.h.s of Eqs. (\ref{eq:dqiscond0}) are not mutually orthogonal,
clearly  there  is  no  $U^\dag_{\bf  C,M}$  such  that  the  recovery
condition (\ref{eq:complementary})  is satisfied.  It  follows that if
Alice  teleports the  state $|0\rangle$  across the  channel  given by
$|\Psi_L\rangle$ in Eq. (\ref{eq:encode}) with the code words given by
Eq. (\ref{eq:ghz}), then under  this configuration the state recovered
is not the secret but simply $|0\rangle$.

\subsection{Example: DQIS with cluster state space 
\label{sec:dqiscluster1}}

Letting $d=2$ in Eq.  (\ref{eq:encode}), we choose $|G_0\rangle$ to be
$|\phi_{0000}\rangle  \equiv |\phi_4\rangle$  in  Eq.  (\ref{eq:lc4}),
and       $|G_1\rangle        \equiv       |\phi_{0101}\rangle       =
Z_2Z_4|\phi_{0000}\rangle$,   with    fiducial   input   state   being
$|0\rangle$. Alice holds the input $|0\rangle$ and qubit 1, Bob qubits
2 and 3, and Rex qubit 4. The secret $|\Psi\rangle = \alpha|0\rangle +
\beta|1\rangle$ is encoded as:
\begin{equation}
|\Psi_L\rangle = \alpha|\phi_{0000}\rangle + \beta|\phi_{0101}\rangle,
\end{equation}
where $|\alpha|^2  + |\beta|^2=1$.  Alice  measures in the  Bell basis
$\{|\Phi^{\pm}\rangle    \equiv    \frac{1}{\sqrt{2}}(|00\rangle   \pm
|11\rangle),  |\Psi^{\pm}\rangle  \equiv \frac{1}{\sqrt{2}}(|01\rangle
\pm   |10\rangle)\}$,   while   Bob   in   the   computational   basis
$\{|00\rangle, |01\rangle,  |10\rangle, |11\rangle\}$, and  Rex in the
$X$ basis.  It may be  verified that these $|G_j\rangle$'s satisfy the
divergence conditions for the above configuration \textbf{C}.  Alice's
outcomes are tabulated in Table \ref{tab:ClusterX}, and Bob's outcomes
corresponding   to   Alice's   outcome   $|\Phi^+\rangle$   in   Table
\ref{tab:ClusterY}.   Based on  Alice's  1  bit and  and  Bob's 2  bit
classical  communication   about  their  outcomes,   Rex  reconstructs
$|\Psi\rangle$.

\begin{widetext}
\begin{center}
\begin{table}[h]
\begin{tabular}{l|l}
\hline Alice's measurement &  State obtained\\ \hline 
$|\Phi^{\pm}\rangle$ & $(\alpha + \beta)(|0+0\rangle + |1+1\rangle)  +  
                  (\alpha - \beta)(|0-1\rangle + |1-0\rangle)$
\\ 
$|\Psi^{\pm}\rangle$    & $(\alpha + \beta)(|0+0\rangle - |1-0\rangle)  +  
                  (\alpha - \beta)(|0-1\rangle - |1+1\rangle)  $ \\
\hline
\end{tabular}
\caption{4-qubit  cluster  state  DQIS:  Alice's measurement  and  the
  (unnormalized)  state obtained  by Bob  and Rex.   Ket $|0+0\rangle$
  represents the  3-qubit state $|0\rangle|+\rangle|0\rangle$,  and so
  on.}
\label{tab:ClusterX}
\end{table}
\end{center}
\end{widetext}

\begin{table}[h]
\begin{tabular}{l|l}
\hline
Bob's measurement & State obtained \\
\hline 
$|00\rangle$ & $ \alpha|+\rangle + \beta|-\rangle$ \\
$|01\rangle$ & $ \alpha|-\rangle + \beta|+\rangle$ \\
$|10\rangle$ & $ \alpha|+\rangle - \beta|-\rangle$ \\
$|11\rangle$ & $ \beta|+\rangle -\alpha|-\rangle $ \\
\hline
\end{tabular}
\caption{4-qubit cluster  state DQIS  (type 1): Bob's  measurement and
  state obtained by Rex.}
\label{tab:ClusterY}
\end{table}

\subsection{Example: DQIS with a QEC code space}

Quantum  error correcting (QEC)  codes \cite{cs,steane}  are $n$-qubit
graph  states  up  to  local  transformations.  A  QEC  code  word  is
stabilized by $n-k$ independent stabilizer operators, where $k$ is the
code rate.

Let $|G_0\rangle$ and $|G_1\rangle$ be respectively, the 5-qubit 1-bit
error   correcting   code  words   introduced   by   Bennett  et   al.
\cite{bensmo}:
\begin{widetext}
\begin{eqnarray}
|0_L\rangle  &=&   \frac{1}{4}  (-  |00000\rangle   -  |11000\rangle  -
|01100\rangle  -  |00110\rangle   -|00011\rangle  -  |10001\rangle  +
|10010\rangle  +  |10100\rangle   +  |01001\rangle  \nonumber  \\  &+&
|01010\rangle  +  |00101\rangle  +  |11110\rangle +  |11101\rangle  +
|11011\rangle    +   |10111\rangle    +    |01111\rangle)   \nonumber\\
|1_L\rangle &=& XXXXX|G_0\rangle,
\label{eq:steane}
\end{eqnarray}
\end{widetext}
where $XXXXX$ signifies  an application of $X$ on  each qubit.  We let
Alice  have qubit  1,  Bob qubits  2  and 3,  Charlie  qubit 4,  while
recoverer Rex  have qubit 5.   We let $|G_0\rangle =  |O_L\rangle$ and
$|G_1\rangle = |1_L\rangle$in Eq.  (\ref{eq:encode}).  Alice teleports
state $|0\rangle$  by measuring in  the Bell basis an  ancillary qubit
prepared   in  that   state,  and   her  part   of   the  entanglement
$|\Psi_L\rangle$ in Eq.  (\ref{eq:encode}). Bob and Charlie measure in
the computational basis. It may be verified that this choice satisfies
the teleportation  divergence condition for the  choice of \textbf{C}.
Alice's,  Bob's, Charlie's  and recoverer  Rex's measurement  data are
tabulated  below in  Tables \ref{tab:VinShor}  and \ref{tab:VinShor1}.
Rex  recovers the  secret based  on  Alice's 1  bit, Bob's  2 bit  and
Charlie's 1  bit classical communication.  Charlie  measures his qubit
in the computational basis $\{|0\rangle,|1\rangle\}$ while Rex applies
the nescessary operation to obtain the qubit.

\begin{widetext}
\begin{center}
{\tiny
\begin{table}[h]
\begin{tabular}{l|l}
\hline
Alice's measurement & State obtained\\
\hline
$|\Phi^{\pm}\rangle $ & $ \alpha(-|0000\rangle -|1100\rangle - |0110\rangle - |0011\rangle
+ |1001\rangle + |1010\rangle + |0101\rangle + |1111\rangle)$ \\
~ & $+ \beta(-|0111\rangle
- |1110\rangle + |1101\rangle + |1011\rangle + |0001\rangle + |0010\rangle + |0100\rangle
+ |1000\rangle)$ \\
\hline
$|\Psi^{\pm}\rangle $ & $ \alpha(-|1000\rangle - |0001\rangle + |0010\rangle + |0100\rangle
+ |1110\rangle + |1101\rangle + |1011\rangle + |0111\rangle)$ \\
~ & $+ \beta(|0110\rangle -
|1111\rangle - |0011\rangle - |1001\rangle - 
|1100\rangle + |0101\rangle + |1010\rangle
+ |0000\rangle)$\\
\hline
\end{tabular}
\caption{Alice's measurement  and state  obtained by Bob,  Charlie and
  Rex in case  of teleportation using the Bennett  et al. 5-qubit code
  (\ref{eq:steane}).}
\label{tab:VinShor}
\end{table}}
\end{center}
\end{widetext}

{\tiny
\begin{table}[h]
\begin{tabular}{l|l}
\hline
Bob's measurement & State obtained \\
\hline 
$|00\rangle$ & $\alpha(-|00\rangle - |11\rangle) + \beta(|01\rangle + |10\rangle)$ \\
$|11\rangle$ & $\alpha(-|00\rangle + |11\rangle) + \beta(|01\rangle - |10\rangle)$ \\
$|01\rangle$ & $\alpha( |01\rangle - |10\rangle) + \beta(|00\rangle - |11\rangle)$ \\
$|10\rangle$ & $\alpha( |01\rangle + |10\rangle) + \beta(|00\rangle + |11\rangle)$ \\
\hline
\end{tabular}
\caption{Bob's measurement and state obtained by rest}
\label{tab:VinShor1}
\end{table}}

\section{Nonlocal subspaces \label{sec:nonlocal}}

We  call two  or more  $n$-qubit  elements $|G_j\rangle$  of a  subset
$\mathcal{D}$ of the graph state basis as degenerate graph states with
respect to a Hermitian operator $P$, if
\begin{equation}
P|G_j\rangle = e|G_j\rangle \forall j \in \mathcal{D},
\label{eq:degenerate}
\end{equation}
where $e$ is a real number.  If $P$ is the Bell operator $\mathcal{B}$
in Eq.  (\ref{eq:bell}), then the  states in $\mathcal{D}$  are called
Bell-degenerate,   and  the   space  spanned   by  the   operators  in
$\mathcal{D}$ as Bell-degenerate graph subspace.

The  set  of all  local-realist  (LR) models  for  a  given number  of
settings  of  Alice,  Bob  et  al.   is  a  polytope  in  a  space  of
correlations. Equations of the  kind (\ref{eq:bell}) correspond to its
facets  \cite{bellperes, fine82}.   Greenberger,  Horne and  Zeilinger
(GHZ) first showed how  entangled states with perfect correlation lead
to  a  dramatic  contradiction  with  LR  models  \cite{ghz}.   Mermin
\cite{bellmermin} pointed  out how  these perfect correlations  can be
used   to  construct   a  Bell-type   inequality  of   the   form  Eq.
(\ref{eq:bell}).  The  problem of deriving  Bell-type inequalities for
various kinds of  graph states has been explored  by different authors
in  a   number  of  directions   \cite{bellmermin,mard,sca05,  genard,
  bellgraph, bellcab, geza, cabguh, vinper, hsugraph, guhcab}.

We denote  by the DQIS  \textit{code basis} $\mathcal{C}$, the  set of
graph basis elements $|G_j\rangle$ in Eq.  (\ref{eq:encode}).  Our aim
is to  construct a Bell  operator with respect  to which all  and only
elements  of $\mathcal{C}$ are  degenerate.  Furthermore,  they should
each  violate  the  corresponding  Bell inequality  to  its  algebraic
maximum of $m$, thereby making $\mathcal{C}$ maximally nonlocal.  This
Bell operator will thus serve as  a witness for the nonlocality of the
encoded  state.   We  present  two complementary  approaches  to  this
problem, discussed in the following two subsections.

\subsection{Bell-degeneracy through unresolvability of
generators \label{sec:degclu}}

Suppose that all $n$ generators $g_j$ of a given $n$-qubit graph basis
are  involved in the  $m$ operators  $h_k$'s that  appear in  the Bell
inequality (\ref{eq:bell}). At most $n$ of them can be independent. If
fewer than $n$  are independent, this gives rise  to other graph basis
state(s) $|G^\prime\rangle$ than  $|G\rangle$ that are consistent with
$\langle\mathcal{B}\rangle  =  m$, and  thus  serves  as  a basis  for
degeneracy.

\begin{thm}
Given an  $n$-qubit graph state $|G\rangle$ that  maximally violates a
Bell  inequality  Eq.    (\ref{eq:bell}),  where  $\mathcal{B}$  is  a
non-trivial  functional   of  all  $n$  $g_j$'s,  if   the  number  of
independent operators  $h_j$ in $\mathcal{B}$  is $r$ ($\le  n$), then
the dimension  of the Bell-degenerate  subspace containing $|G\rangle$
is $2^{n-r}$.
\label{thm:1}
\end{thm}
\textbf{Proof.}   The   maximal  violation  of   the  Bell  inequality
(\ref{eq:bell}) by  $|G\rangle$ implies that this  state satisfies the
$m$  constraints $\forall_{j=1}^m h_j  \rightarrow +1$.   If $r  = n$,
then one can solve for the  $g_j$'s to obtain a unique solution, which
must be $\forall_{j=1}^n g_j  = +1$, corresponding to $|G\rangle$.  If
$r < n$, then there are fewer constraints than variables ($n$).  Since
the  $g_j$ are  two-valued ($\pm  1$), this  corresponds  to $2^{n-r}$
possible  $g_j$-assignments  consistent   with  the  $m$  constraints.
\hfill $\blacksquare$
\bigskip

If fewer than $n$ generators  appear in $\mathcal{B}$, then there will
be additional degeneracy, by virtue of Theorem \ref{thm:2} below.  Let
$\Xi$ denote  the set of  Bell-degenerate graph basis states.   If the
code  rate of  a  DQIS protocol  is  $k$ bits,  we  must choose  $2^k$
elements  from $\Xi$ that  satisfy the  conditions (\ref{eq:dqiscond})
and  (\ref{eq:complementary}).  A  necessary  condition for  this  is,
clearly, $r+k \le n$.

An  example: The  linear  cluster state  $|\phi_{0000}\rangle$ in  Eq.
(\ref{eq:lc4})  is  a graph  state  corresponding  to the  stabilizing
operators $g_1  = XZII \rightarrow  +1$, $g_2 \equiv  ZXZI \rightarrow
+1$, $g_3 \equiv IZXZ \rightarrow +1$ and $g_4 \equiv IIZX \rightarrow
+1$.  The Bell operator
\begin{eqnarray}
\mathcal{B}^{\phi}_1 &=& h_1 + h_2 + h_3 + h_4 \nonumber \\
  &=&g_1g_3 + g_2g_3 + g_1g_3g_4 + g_2g_3g_4 \nonumber \\
&=& XIXZ + XIYY + ZYYZ - ZYXY,
\label{eq:sca05bell}
\end{eqnarray}
for which $q=3$ (at most  only 3 terms in Eq. (\ref{eq:sca05bell}) can
be  simultaneously  made  positive  when assigned  determinate  values
$\pm1$ non-contextually),  so that the  local-realist bound is  2.  It
attains  the algebraically  maximum  value of  $m=4$  when applied  to
$|\phi_{0000}\rangle$ \cite{sca05}.   The product of any  three of the
four summands  in Eq.  (\ref{eq:sca05bell}) is equal  to the remaining
one,  implying  that the  4  constraints $\forall_j  h_j\rightarrow+1$
imposed  by these  operators  are  not independent;  only  3 are.   By
Theorem \ref{thm:1}, this corresponds to a Bell-degenerate subspace of
dimension 2.  Solving  for the $g_j$'s we find $g_4  = h_1h_2 = h_3h_4
\rightarrow +1$.   The remaining three generators  cannot be resolved,
but are  subject to  the condition $g_1g_2  = h_1h_3  \rightarrow +1$,
$g_2g_3  =  h_1h_2h_3  \rightarrow   +1$,  which  is  consistent  with
$g_1=g_2=g_3=-1$,  apart from  of course  $g_1=g_2=g_3=+1$.   Thus, we
find that $\Xi$ additionally contains the state with the \textit{graph
  signature} $(-1,-1,-1,1)$, which corresponds to the state
\begin{eqnarray}
|\phi_{1110}\rangle        &=&       Z_1Z_2Z_3|\phi_{0000}\rangle   
\nonumber \\    &=&
  \frac{1}{2}\left(|{-}0{-}0\rangle      +      |{-}0{+}1\rangle     -
  |{+}1{+}0\rangle - |{+}1{-}1\rangle\right),\nonumber \\
\label{eq:phi1110}
\end{eqnarray}
which also  yields $\langle \mathcal{B}\rangle =  4$.  A teleportation
configuration   under  which   these   two  elements   of  $\Xi$   are
teleportation-divergent is thus suitable for secure DQIS.

Suppose $l$ stabilizer  generators $f_j$ appear in all  the $h_j$'s in
Eq. (\ref{eq:bell}). If $|G\rangle$ is a state that maximally violates
Eq.  (\ref{eq:bell}), then the  $l-1$ constraints imposed by the above
requirement correspond to  $\sum^{l}_{j =0; j\textrm{~even}} {^rC}_j =
2^{l-1}$ states  obtained by  flipping the sign  of an even  number of
these   generators,   since  they   preserve   the  value   assignment
$f_1f_2\cdots  f_l \rightarrow  +1$, and  hence the  value assignments
$h_j \rightarrow +1$ on these states. For the Bell operator
\begin{eqnarray}
\mathcal{B}^{\phi}_2 &=& g_2g_4(1 + g_1)(1 +  g_3) \nonumber \\
&=& ZXIX + YYIX + YXXY
- ZYXY,
\label{eq:sca05bell1}
\end{eqnarray}
which  also has the  local-realist bound  of 2,  and the  same quantum
bound  of 4  on the  state $|\phi_{0000}\rangle$  \cite{guhcab}.  Only
three of the four summands  in the rhs of Eq. (\ref{eq:sca05bell}) are
independent,  so that  by Theorem  \ref{thm:1}, the  dimension  of the
Bell-degenerate subspace $2$, which  is immediately seen to correspond
to the  state obtained by flipping  the sign of both  $g_2$ and $g_4$:
i.e.,  the  state  with   the  graph  signature  $(1,-1,1,-1)$,  which
corresponds  to  the 4-qubit  cluster  state $|\phi_{0101}\rangle$  of
Section \ref{sec:dqiscluster1}.

\subsection{Bell-degeneracy via a stabilized subspace \label{sec:qecc}}

The other  method is applicable  to $(n,k)$ \textit{graph  codes}, the
subspace  of  a  $n$-qubit  states,  stabilized  by  $n-k$  stabilizer
generators $g_j$, with $k > 0$.  We denote by $\mathcal{P}$ the set of
these  generators.  Let  us  denote by  $\mathcal{S}|_{\cal  P}$,  the
restriction of $\mathcal{S}$ to products of elements in $\mathcal{P}$.
Up to local transformations, QEC codes are graph codes.
 
\begin{thm}
Given a $(n,k)$ graph code $\mathcal{G}$ stabilized by operators $g_j$
($1 \le  j \le n-k$),  any Bell operator  $\mathcal{B}_\mathcal{P}$ of
the type  (\ref{eq:bell}), obtained by  adding only elements  $h_j \in
\mathcal{S}|_{\mathcal{P}}$  induces  a  Bell-degenerate  subspace  of
dimension $\ge 2^k$.  If $n-k$  of these $h_j$'s are independent, then
the dimension is exactly $2^k$.
\label{thm:2}
\end{thm}
\textbf{Proof.}    For   each  of   the   $2^k$   graph  basis   state
$|G^\prime\rangle$  stabilized by  $g_j$'s  in $\mathcal{P}$,  clearly
$h_j  \rightarrow  +1$   for  summands  in  $\mathcal{B}_\mathcal{P}$,
implying that the Bell inequality  is maximally violated for any state
in  the subspace  spanned by  these basis  states.  If  the  number of
independent stabilizers  in $\mathcal{B}_\mathcal{P}$ is  $r \le n-k$,
then by virtue of Theorem \ref{thm:1}, the dimension of the degenerate
subspace is $2^k \times 2^{(n-k)-r}  = 2^{n-r}$.  Setting $r=n-k$ in the
last  expression  above,  we   obtain  $2^{k}$,  as  desired.   \hfill
$\blacksquare$
\bigskip


QECC code states  are graph states, that satisfy  the graph conditions
(\ref{eq:graph}) up to local transformations. A set of four stabilizer
operators (which need  not have the error correcting  property for our
purpose) for the above 5-qubit code (\ref{eq:steane}) are:
\begin{eqnarray}
g_1 &=& XYYXI \nonumber \\
g_2 &=& IXYYX \nonumber \\
g_3 &=& ZYIYZ \nonumber \\
g_4 &=& XYZYX.
\label{eq:bellgenVinShor}
\end{eqnarray}
from which we  can construct the following stabilizers,  which we cast
in the form of a GHZ contradiction with local realism:
\begin{eqnarray}
h_1 &\equiv & g_1g_3g_4  =  ZYXXY \longrightarrow +1 \nonumber \\
h_2 &\equiv& g_1g_4     =  -IIXZX \longrightarrow -1 \nonumber \\
h_3 &\equiv& g_2g_3     =  ZZYIY \longrightarrow +1 \nonumber \\
h_4 &\equiv& g_1g_2     =  XZIZX \longrightarrow +1 \nonumber \\
h_5 &\equiv& g_1 =  XYYXI \longrightarrow +1.
\label{eq:bellopVinShor1}
\end{eqnarray}
This constitutes a GHZ contradiction \cite{ghz} with any local-realist
assignment of  definite values to $X, Y,  Z$ as can be  as follows: in
the operators  $h_j$, a  Pauli operator always  appears twice  along a
vertical column,  implying that the  product of the $h_j$'s  should be
1.  Yet  the product  of  the above  value  assignments  to the  state
$|G\rangle$ is $-1$. This logical  contradiction means that the $X, Y,
Z$'s of the  particles cannot be thought of  as possessing determinate
values independent of the  measurement context (the choice of settings
on the other particles).

From  Eq.  (\ref{eq:bellopVinShor1})  one  can  write  down  the  Bell
operator
\begin{subequations}
\begin{eqnarray}
\mathcal{B}_{5q} &=& h_1 + h_2 + h_3 + h_4 + h_5, \label{eq:aa} \\ &=&
g_1g_4(g_3 + 1) + g_2(g_3+g_1) + g_1. \label{eq:bb}
\label{eq:bellsteane}
\end{eqnarray}
\end{subequations}
Eq. (\ref{eq:bellsteane}) satisfies the Bell inequality
\begin{equation}
\langle \mathcal{B}_{5q} \rangle \le 3,
\label{eq:bellboundq5}
\end{equation}
as can be seen by evaluating $\mathcal{B}_{5q}$ for all possible $2^5$
value  assignments $\pm1$  to $X,  Y, Z$'s  in Eq.   (\ref{eq:aa}). By
virtue  of  Theorem \ref{thm:2},  both  the  states $|0_L\rangle$  and
$|1_L\rangle$  of  (\ref{eq:steane})  as  well  as  any  superposition
thereof,  violate the  above  inequality maximally,  by  the value  5.
There  is no  further degeneracy,  since four  of the  $h_j$'s  in Eq.
(\ref{eq:bellopVinShor1}) are independent.  Solving for the $g_j$'s we
obtain: $g_1=h_5; g_2 = h_4h_5; g_3 = h_3h_4h_5$ and $g_4 = h_2h_5$.

Here  are two  examples  where both  the  above two  theorems must  be
invoked.  The stabiliziers for  the Steane \cite{steane} code are $g_1
= X_4 X_5 X_6 X_7, g_2 = X_2 X_3 X_6 X_7, g_3 = X_1 X_3 X_5 X_7, g_4 =
Z_4 Z_5 Z_6 Z_7,  g_5 = Z_2 Z_3 Z_6 Z_7$ and $g_6  = Z_1 Z_3 Z_5 Z_7$,
with a Bell inequality taking the form
\begin{eqnarray}
\langle \mathcal{B}_{\rm Steane} \rangle &=&
 \langle h_1 + h_2 + h_3 + h_4 + h_5 + h_6 \rangle \nonumber \\ 
 &=& \langle g_1g_2(g_4 + g_4g_5
+ 1) + g_3g_5(g_2 + g_1) + g_5\rangle \nonumber \\
 &\le& 4,
\label{eq:bellboundst}
\end{eqnarray}
where $h_1  \equiv g_1g_2g_4$, $h_2 \equiv  g_1g_2g_4g_5$, $h_3 \equiv
g_1g_2$, and  so on sequentially.  The quantum  mechanical case yields
the maximal $\langle  \mathcal{B} \rangle = 6$ (the  number of $h_j$'s
in  Eq.  (\ref{eq:bellboundst}))  for any  state  $\alpha|0_L\rangle +
\beta|1_L\rangle$ in the code space  of the Steane code.  In this case
$\mathcal{P} =  \{g_1,g_2,g_3,g_4,g_5\}$ implying that  all $2^{7-5} =
4$  7-qubit  basis states  stabilized  by  these  5 operators  span  a
Bell-degenerate subspace.

In  addition,   only  4  of   the  $h_j$'s  are  independent   in  Eq.
(\ref{eq:bellboundst}), in that  the following two constraints $h_1h_2
= h_6$  and $h_4h_5 =  h_3$ appear. Thus  for any value  assignment of
$g_6$  and $g_7$,  there are  $2^{5-4} =  2$ graph  basis  states that
maximally violate Eq.  (\ref{eq:bellboundst}), which are determined by
solving  for  the 5  $g_j$'s  in terms  of  the  6 operators  $h_j$'s.
Solving,  we find  $g_4 =  h_2h_3h_6 \rightarrow  +1$ and  $g_5  = h_6
\rightarrow +1$, while $g_1 g_2 =  h_4 \rightarrow +1$, $g_1 g_3 = h_3
h_5  \rightarrow +1$  and $g_2  g_3  = h_2h_5  \rightarrow +1$.   This
corresponds  to   the  graph  signatures   $(1,  1,  1,  1,   1)$  and
$(-1,-1,-1,1,1)$  in the  first five  $g_j$'s.  Thus  in all  there is
Bell-degeneracy of $4 \times 2 = 8$.

The  stabiliziers  for  the  Shor  code  \cite{shorcode}  are  $g_1  =
ZZIIIIIII, g_2  = IZZIIIIII,  g_3 = IIIZZIIII,  g_4 = IIIIZZIII,  g_5 =
IIIIIIZZI,  g_6 =  IIIIIIIZZ, g_7  = XXXXXXIII,  g_8 =  IIIXXXXXX$ for
which a Bell inequality takes the form
\begin{eqnarray}
\langle \mathcal{B}_{\rm Shor}  \rangle &\equiv& 
 \langle h_1 + h_2 + h_3 + h_4 + h_5 + h_6 + h_7 \rangle \nonumber \\ 
&=& \langle g_3g_8(g_1g_4 +
g_5g_7 + g_2 + g_7) + g_8(g_4g_5 + 1) \nonumber \\
&+&  g_1g_2 \rangle \le 5,
\label{eq:bellShor}
\end{eqnarray}
where  $h_1  \equiv  g_3g_8g_1g_4$,  $h_2 \equiv  g_3g_8g_5g_7$,  $h_3
\equiv  g_3g_8g_2$,  and  so  on,  sequentially.   while  the  quantum
mechanical  state yields  $\langle \mathcal{B}  \rangle =  7$  for any
state $\alpha|0_L\rangle + \beta|1_L\rangle$  in the code space of the
Shor      code.       In       this      case      $\mathcal{P}      =
\{g_1,g_2,g_3,g_4,g_5,g_7,g_8\}$ implying that  all four 9-qubit basis
states  stabilized  by  these   7  operators  span  a  Bell-degenerate
subspace.

In  addition,  only  6  of  the  7  $h_j$'s  are  independent  in  Eq.
(\ref{eq:bellboundst}),   giving  rise   to  2   degenerate   sets  in
$\mathcal{P}$.  Thus for any of  four value assignments to $g_6, g_9$,
there  are   two  graph  basis  states  that   maximally  violate  Eq.
(\ref{eq:bellShor}), which are determined by solving for the 7 $g_j$'s
in $\mathcal{P}$ in terms of the $h_j$'s. This yields $g_4 = h_1h_3h_7
\rightarrow +1$, $g_5 = h_1h_3h_5h_6h_7 \rightarrow +1$ and $g_8 = h_6
\rightarrow +1$,  while $g_1 g_3 = h_1h_2h_4h_5  \rightarrow +1$, $g_2
g_3 = h_3 h_7 \rightarrow +1$,  $g_1g_2 = h_6 \rightarrow +1$ and $g_3
g_7  =  h_4h_7  \rightarrow   +1$.   This  corresponds  to  the  graph
signatures $(1, 1, 1, 1, 1, 1,  1)$ and $(-1, -1, -1, +1, +1, -1, +1)$
for  the   $g_j$'s  in  $\mathcal{P}$.    Thus,  in  all,   we  obtain
Bell-degeneracy of $4 \times 2 = 8$.

\section{Security consideration: towards a device-independent scenario \label{sec:security}}

Although  the security  of quantum  key distribution  (QKD),  has been
known for  some time, recent  work has uncovered the  close connection
between security (that legitimate participants can distil secret bits)
and the violation of a Bell  inequality, both in the two-party as well
as  multi-party \cite{sg1,sg2}  scenarios, an  intuition  that already
exists in the Ekert protocol \cite{ekert}.

This connection  has assumed further  importance for other  reasons: a
proof of  security based on the  violation of Bell  type inequality is
expected  to   hold  good   in  any  nonlocal,   non-signaling  theory
\cite{diqkd0},   and   even   in   a   device   independent   scenario
\cite{diqkd2},  i.e.,   one  where  there   is  a  lack   of  complete
characterization of  devices used. The eavesdropper may  be the vendor
from whom Alice and Bob purchase their (entangled) states and devices.
Eve  may  insert  hidden  dimensions  into the  devices,  and  unknown
correlations into  the states, that  would empower side  channels that
leak to  her information about  Alice's and Bob's  measurement choices
and outcomes.  

Thus a  conventional check of error  rates will not  do.  The legimate
parties must verify that the correlation data has not been produced by
a  separable  state  in  a  larger  dimensional  space  \cite{diqkd1}.
However,  if  the legitimate  parties  verify  via simultaneous  local
operations  and  classical communication  that  a  Bell inequality  is
violated to  a sufficiently  high level, then,  assuming no-signaling,
the   correlations  are  guaranteed   to  allow   distillable  secrecy
\cite{bell2sqkd,sxnosigcor}.   This is a  consequence of  the monogamy
\cite{cof00}  of  quantum correlations  and  holds  good in  nonlocal,
non-signaling  theories  \cite{mag06},   though  no-signaling  is  not
necessary \cite{pawmon}.

Let us consider  the 4-qubit cluster state DQIS  considered in Section
\ref{sec:dqiscluster1}  applied to the  protocol described  earlier in
Section  \ref{sec:dqis}.  Alice  holds  qubit  1, Bob  2  and 3,  and,
finally, Rex qubit 4 of $N$ copies of the encoded version of the state
$|\Psi\rangle$  or  one of  its  encrypted  versions.  Alice  randomly
announces the  serial number of one  of these copies.  On  each of the
remaining $N-1$ copies, Alice  randomly makes a measurement drawn from
the set $S_A = \{Y, Z\}$, Bob from the set $S_B = \{XI, YI, XX, YX\}$,
and Rex  from $S_R  = \{X, Y\}$.   They classically  communicate their
measurements and outcomes to Alice.   About $4/(2 \times 4 \times 2) =
1/4$  of these  are found  to have  measured one  of the  four 4-qubit
observables  appearing as  a  summand in  $\mathcal{B}^\phi_2$ in  Eq.
(\ref{eq:sca05bell1}), and  the parties  verify that the  outcomes are
consistent with the sufficiently high violation of the Bell inequality
$\langle  \mathcal{B}^\phi_2\rangle  \le  2$.  Importantly,  by  prior
synchronization of  clocks, each participant  must measure her  or his
observable  simultaneously, in  order to  avoid the  possibility  of a
signaling   from   the    source   to   the   measurement   appartuses
\cite{esther}. Bounding  the timing of the  classical communication is
used to ensure this.

Conditioned on their passing the  Bell test, they proceed to implement
the   DQIS  protocol   to   allow  Rex   to   reconstruct  the   state
$|\Psi\rangle$.  If  not, then  they may either  abort the run  of the
protocol, and restart from a fresh distribution of entangled states.

As an illustration of the role of monogamy, we consider below a simple
single-qubit attack  by Eve  on a DQIS  protocol based on  the 5-qubit
state (\ref{eq:steane}), which produces a lowering of the violation of
the Bell inequality observed by  Alice, Bob, and Rex.  This is because
an attempt  by Eve  to extract information  entangles her  system with
theirs,   causing  the  latter   to  diffuse   from  the   code  space
span$\{|G_0\rangle,   |G_1\rangle\}$.  It   is   important  that   the
degeneracy is restricted to  the code space, since otherwise diffusion
of the state to degenerate non-coding sectors would not be detected by
looking for a reduction in the Bell inequality violation.

In  the  device independent  scenario,  Eve's  hidden dimensions  will
become entangled with the  the legitimate particles, thereby producing
a  detectable  dip in  the  observed  maximal  violation of  the  Bell
inequality  (\ref{eq:bellboundq5}).  Thus sufficiently  high violation
of this inequality guarantees that the entangled state lies within the
code space, and is uncorrelated with unknown degrees of freedom.

Suppose Eve attacks  the fourth qubit of an encoded  state in the span
of the codewords (\ref{eq:steane}), via  a 1-qubit attack given by the
interaction:
\begin{equation}
U(\theta) = \frac{1 + Z}{2}\otimes\mathbb{I} + \frac{1-Z}{2}\otimes
\left( \begin{array}{cc} \cos\theta & \sin\theta \\
\sin\theta & -\cos\theta \end{array}\right),
\end{equation}
which  continuously  varies from  an  identity  operation  to CNOT  as
$\theta$ ranges in $[0,\pi/2]$.

Applying  $U$ on the  encoded state  and her  ancilla prepared  in the
state $|0\rangle$, Eve transforms an arbitrary logical state as:
\begin{widetext}
\begin{eqnarray}
(\alpha|0_L\rangle   +   \beta|1_L\rangle)|0\rangle  &\longrightarrow&
  |\Psi\rangle_{ABCDE}    \nonumber    \\    &=&    \frac{1}{\sqrt{2}}
  \left(\left[\alpha|0_L;0\rangle                                     +
    \beta|1_L;0\rangle\right]|0\rangle_E + \left[\alpha|0_L;1\rangle +
    \beta|1_L;1\rangle\right](C|0\rangle + S|1\rangle) \right),
\label{eq:atak}
\end{eqnarray}
\end{widetext}
where $|j_L;k\rangle$ denotes the  superposition of terms in the above
5-qubit  code word  encoding  bit $j$  having  bit $k$  in the  fourth
position;  $C$  and  $S$  denote  $\cos(\theta)$  and  $\sin(\theta)$,
respectively.   The  reduced  density  operator  for  state  with  the
legitimate agents is given by:
\begin{widetext}
\begin{eqnarray}
\rho_{ABCD}  &=& \frac{1}{2}\left( \left(  \left[\alpha|0_L;0\rangle +
  \beta|1_L;0\rangle    \right]    +   C\left[\alpha|0_L;1\rangle    +
  \beta|1_L;1\rangle \right]  \right) \left( \left[\alpha\langle0_L;0|
  +   \beta\langle1_L;0|   \right]   +  C\left[\alpha\langle0_L;1|   +
  \beta\langle1_L;1|  \right] \right)\right.  \nonumber \\  &+& \left.
S^2\left(    \alpha|0_L;1\rangle    +    \beta|1_L;1\rangle    \right)
\left(\alpha\langle0_L;1| + \beta\langle1_L;1| \right)\right),
\label{eq:sta}
\end{eqnarray}
\end{widetext}
which  has  support  in  the  4-dimension  Hilbert  space  spanned  by
$\{|0_L;0\rangle, |0_L;1\rangle, |1_L;0\rangle, |1_L;1\rangle\}$.

Since   $h_m|j_L\rangle  =  |j_L\rangle$,   it  follows   that  either
$h_m|j_L;k\rangle$          equals          $|j_L;k\rangle$         or
$|j_L;\overline{k}\rangle$.       In     particular,      from     Eq.
(\ref{eq:bellgenVinShor}), it  follows that the action  of the $h_m$'s
is  to leave $|j;k\rangle$  invariant or  to toggle  it in  the second
index (when there is a $X$ or $Y$ in the 4th index). Thus:
\begin{equation}
h_m|j_L;k\rangle = 
\left\{ \begin{array}{cc}
|j_L;k\rangle & ~~(m=1,2,5) \\
 |j_L;\overline{k}\rangle, & ~~(m=3,4) 
\end{array}\right.
\end{equation}
One then finds that
\begin{equation}
\textrm{Tr}_E\left(h_m\rho_{ABCD}\right) =
\left\{
\begin{array}{cc}
\textrm{Tr}_E\left(\rho_{ABCDE}\right) = \pm 1 & ~~(m=2,3,4) \\
-\cos(\theta) & ~~(m=1,5),
\end{array} \right.
\end{equation}
from  which  it  follows  that 
\begin{equation}
\langle \mathcal{B}_{5q} \rangle = 3 + 2\cos(\theta),
\label{eq:monogamyatak}
\end{equation}
implying that the attack can be witnessed by a reduction in the degree
of violation  of the Bell  inequality (\ref{eq:bellboundq5}), dropping
all the  way down the  local-realistic bound of  3 when the  attack is
maximal with $\theta=\pi/2$.  It should be noted in  the general case,
the  tolerable  Bell  inequality  violation  will be  well  above  the
local-realist bound.


\section{Conclusions \label{sec:conclusions}}

The protocol  of DQIS, which inverts  the role of the  input state and
channel, will  be useful  in situations where  the dealer  has bounded
quantum memory, and cannot stock  secrets but is able simply prepare a
fixed state when transmission  is needed.  Classical encryption can be
used  to protect  the encoded  state  if required.   The coding  graph
states must  possess suitable teleportation  divergence properties for
DQIS to work,  and must be Bell degenerate for  proving security via a
Bell test  on the encoded state.  We studied two  methods of producing
Bell degeneracy.   A simple  example of DQIS  with a 5-qubit  QECC was
presented.  The use of the nonlocal properties of the code states is a
useful way  to perform security check,  and particularly indispensible
in the device  independent scenario.  Further, a proof  of security is
expected to hold good in  any non-signaling, nonlocal theory, which is
useful in  the unlikely event that  quantum mechanics turns  out to be
invalid.

\bibliography{axta}

\end{document}